\documentclass[fleqn,usenatbib]{mnras}

\usepackage[T1]{fontenc}
\usepackage{ae,aecompl}

\newcommand\wm{$\omega$}
\newcommand\wcrit{$\omega_{\rm crit}$}

\newcommand\msun{\rm{M$_{\odot}$}}

\newcommand\teff{T$_{\rm eff}$}

\usepackage[usenames,dvipsnames]{color}

\def\simgt{\lower.5ex\hbox{$\; \buildrel > \over \sim \;$}}
\def\simlt{\lower.5ex\hbox{$\; \buildrel < \over \sim \;$}}

\usepackage{graphicx}	
\usepackage{amsmath}	
\usepackage{amssymb}	

\title[Stellar Rotation in the eMSTO cluster NGC 1856]{The extended Main Sequence Turn Off cluster NGC1856:\\ 
rotational evolution in a coeval stellar ensemble}
\author[F. D'Antona et al.]{F. D'Antona$^{1}$, M. Di Criscienzo$^{1}$, T. Decressin$^{1}$, A. P. Milone$^{2}$, E. Vesperini$^{3}$ \newauthor  \& P. Ventura$^{1}$ 
\thanks{E-mail: franca.dantona@gmail.com (FD)}
\\ \\
$^{1}$ INAF, Osservatorio Astronomico di Roma, Via Frascati 33, 
I-00040 Monteporzio Catone (Roma), Italy.\\
$^{2}$  Research School of Astronomy \& Astrophysics, Australian National University, Canberra ACT 2611, Australia\\
$^{3}$ Department of Astronomy, Indiana University, Swain West, 727 E. 3rd Street, IN 47405 Bloomington (USA)\\
}
\date{Accepted XXX. Received YYY; in original form ZZZ}

\pubyear{2015}

\begin{document}
\label{firstpage}
\pagerange{\pageref{firstpage}--\pageref{lastpage}}
\maketitle

\begin{abstract}
Multiple or extended turnoffs in young clusters in the Magellanic Clouds have recently received large attention. A number of studies have shown that they may be interpreted as the result of a significant age spread  (several 10$^8$yr in clusters aged 1--2~Gyr), while others attribute them to a spread in stellar rotation. We focus on the cluster NGC~1856, showing a splitting in the upper part of the main sequence, well visible in the color m$_{\rm F336W}$-m$_{\rm F555W}$, and a very wide turnoff region. Using population synthesis available from the Geneva stellar models, we show that the cluster data can be interpreted as superposition of two main populations having the same age ($\sim$350~Myr), composed for 2/3 of very rapidly rotating stars, defining the upper turnoff region and the redder main sequence, and for 1/3 of slowly/non--rotating stars. Since rapid rotation is a common property of the B-A type stars, the main question raised by this model concerns the origin of the slowly/non-rotating component. Binary synchronization is a possible process behind the slowly/non-rotating population; in this case, many slowly/non-rotating stars should still be part of binary systems with orbital  periods in the range from 4 to  500 days. Such periods imply that Roche lobe overflow occurs, during the evolution of the primary off the main sequence, so most primaries may not be able to ignite core helium burning, consistently why the lack of a red clump progeny of the slowly rotating  population. 
\end{abstract}




\section{Introduction}

\label{Intro}
Extended main sequence turnoff regions (eMSTO) appear to be a typical feature of massive intermediate--age globular clusters (GCs) in both Magellanic Clouds, as shown by deep observations of their color--magnitude diagrams (CMDs) taken with the ACS and WFC3 onboard the Hubble Space Telescope (HST). 
Most of these clusters are in the age range 1--2~Gyr  \citep{mackey2008, milone2009}, and, if the eMSTOs are interpreted as due to differences in the formation epoch of stars, their age spreads range from 150  to $\sim$500~Myr \citep{milone2009, girardi2011, goudfrooij2011, rubele2013} .
\begin{figure*}    
\centering{
\includegraphics[width=6cm]{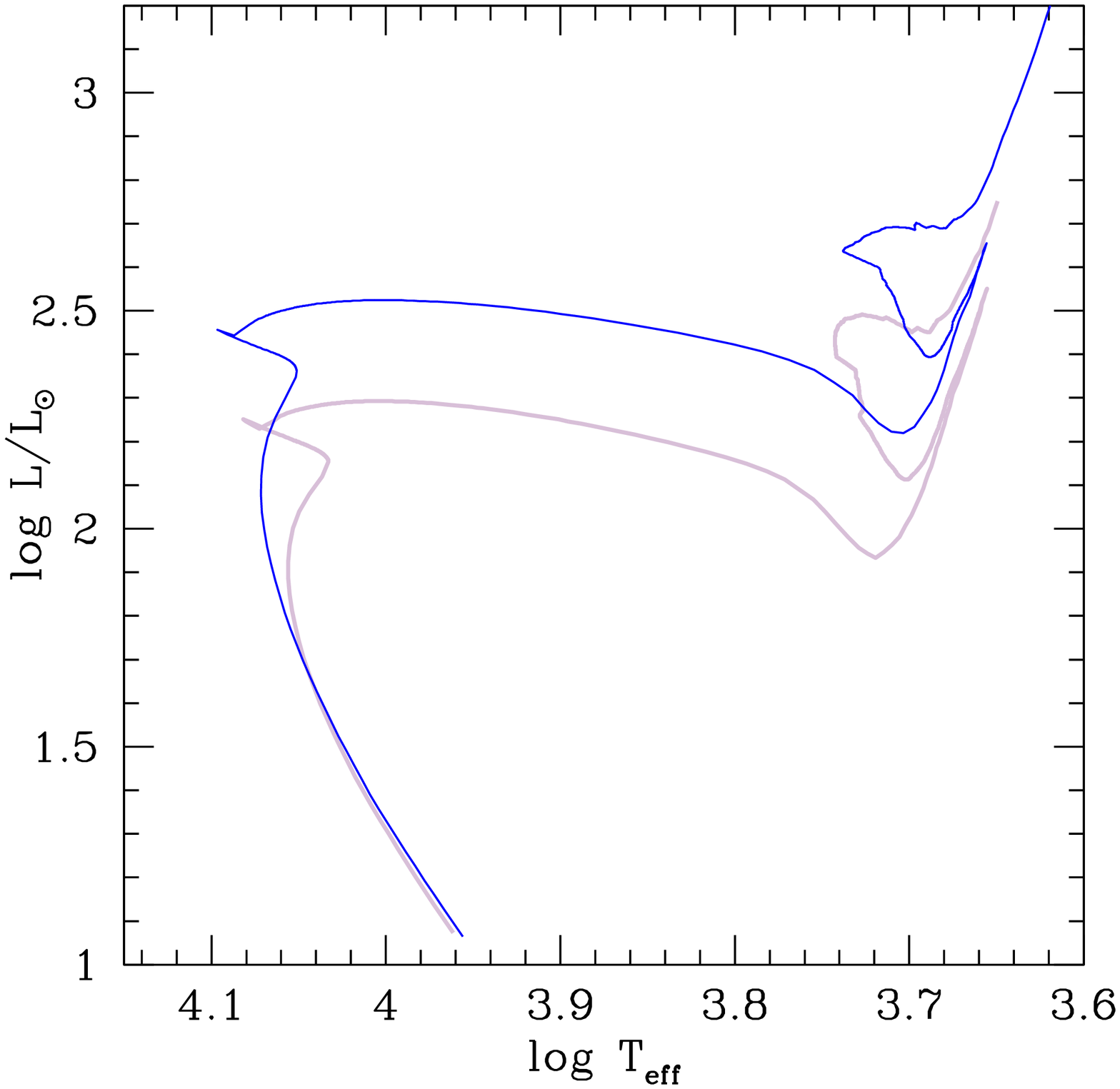}\includegraphics[width=6cm]{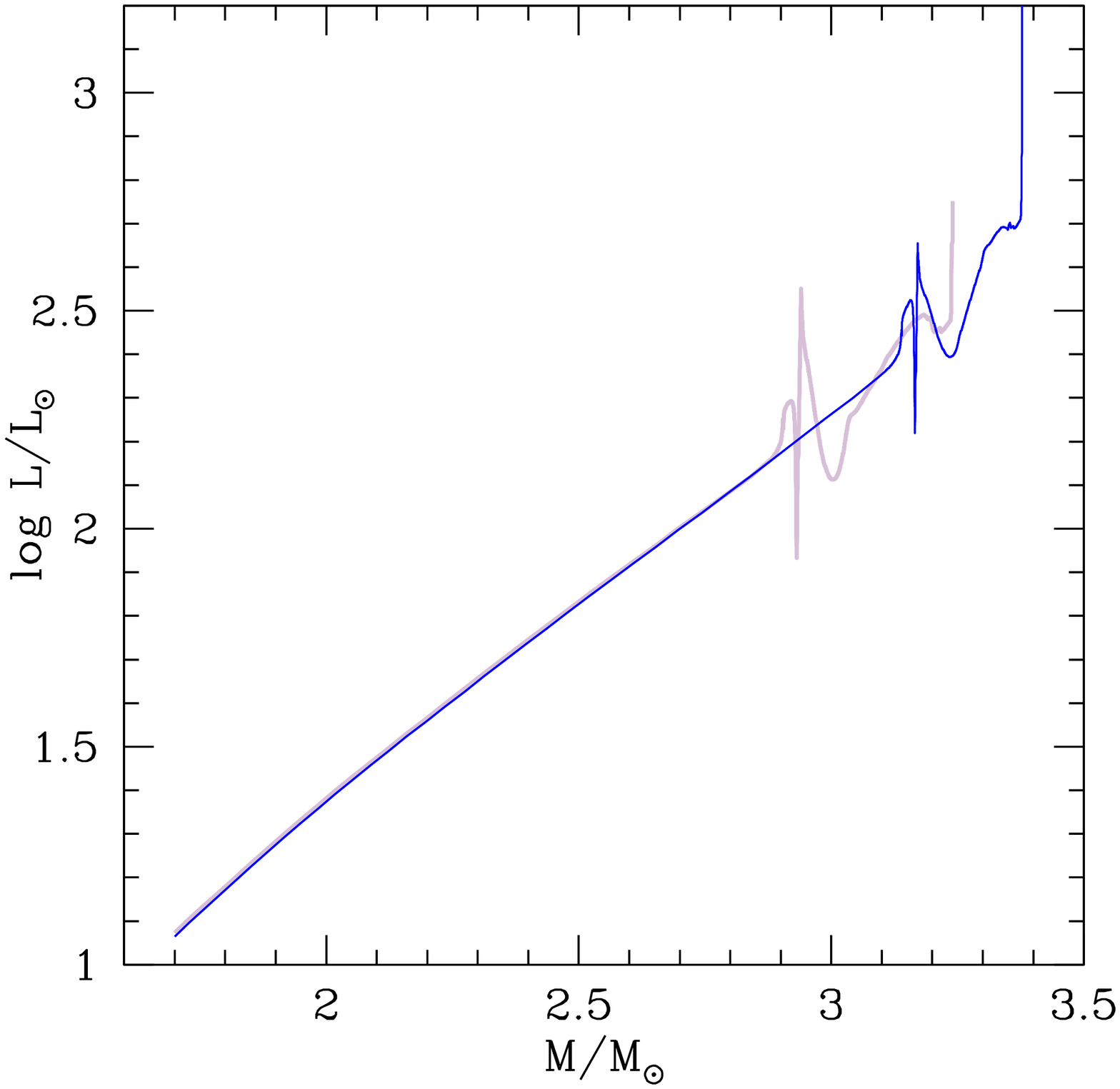}\includegraphics[width=6cm]{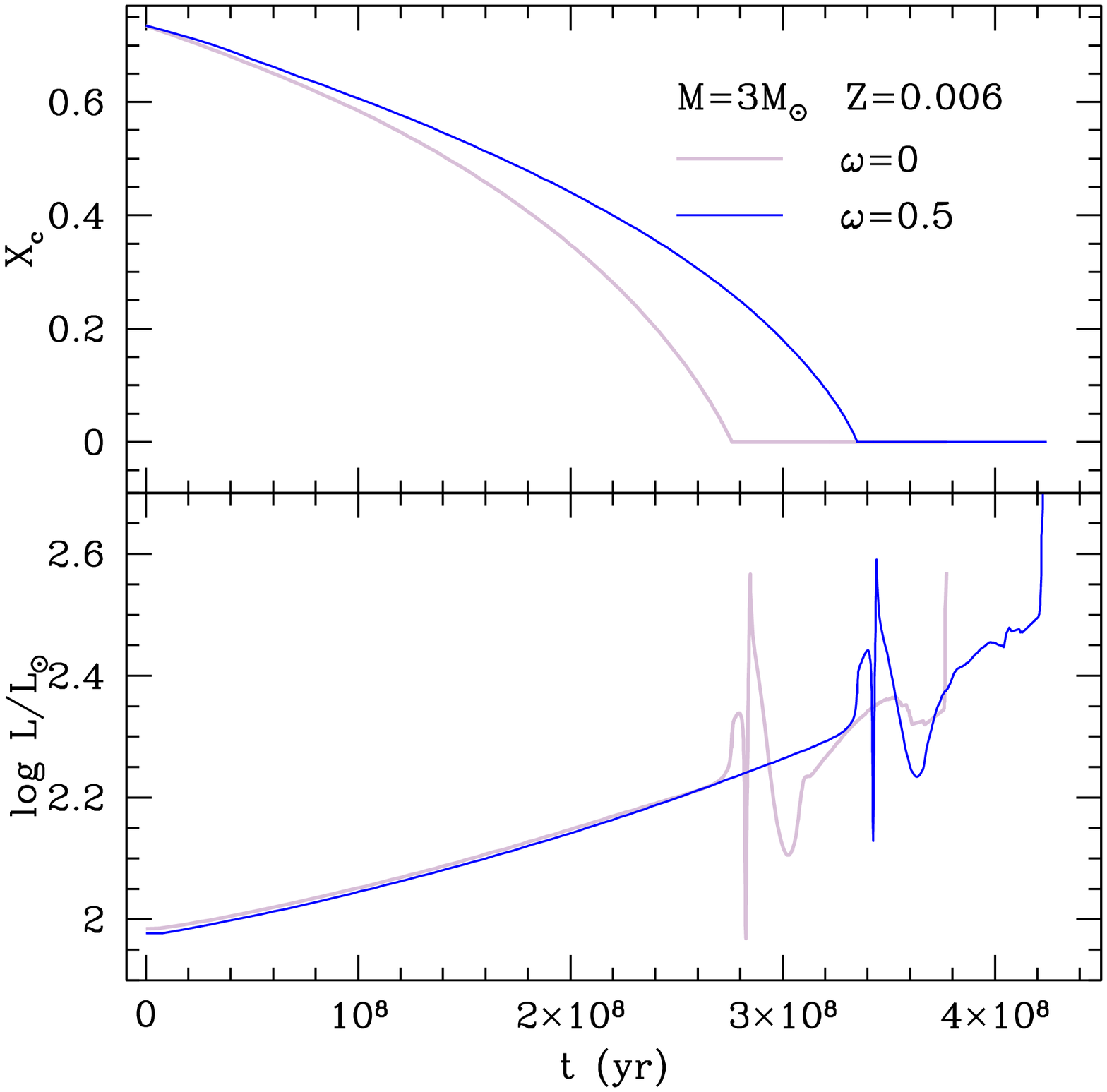}
}
\caption{Left: 300~Myr isochrones for Z=0.006 from Georgy et al. (2013) for non rotating models (grey line) and models rotating at $\omega$=0.5~$\omega_{\rm crit}$\ (blue line); center: luminosity versus mass for the same isochrones; right side: time evolution of the 3\msun\ non-rotating and rotating models.}
\label{f1} 
\end{figure*}
 This possible age spread has been sometimes used  \citep{conroyspergel2011} as a possible evidence of the age difference underlying multiple generations in old galactic GCs \citep{gratton2012, piotto2012}, although theoretical work on these latter shows that the possible age differences must be contained within about 100~Myr \citep{dercole2010}. 
Moreover, (limited) spectroscopic observations of stars in the eMSTO cluster NGC\,1806 seem to suggest that this cluster has homogeneous abundance of O, Na, Al and Mg, 
 and that the O--Na anti correlation may be not present in clusters with extended turnoffs \citep{mucciarelli2014ona}. If confirmed by larger dataset of spectroscopic studies in clusters having eMSTO, the result would indicate the lack of any connection between eMSTO clusters and multiple-populations GCs, since the anticorrelation between oxygen and sodium abundances is one of the key fingerprints of multiple populations in old GCs. Notice that, in galactic GCs, the Na--O anticorrelation is present also at metallicities  as large as those of LMC clusters, in which oxygen is slightly depleted, but the sodium abundance spans $>$0.5~dex \citep{carretta2009}. Other LMC intermediate age clusters may have sodium spreads by 0.1--0.35~dex, which may hint to a possible link with the presence of second generation stars \citep[see the discussion in][]{goudfrooij2014}, but the data \citep{mucciarelli2008} are as limited  as those in NGC~1806, to allow final conclusions.
A lack of the Na--O anticorrelation would call into question any attempt to use observations  of the eMSTO young and intermediate-age clusters, to rule out formation and self-enrichment models for old GCs.

 \cite{bastiandemink2009}  suggested that stellar rotation may cause the eMSTO, as stellar rotation affects the structure of the star and the inclination angle of the star relative to the observer will change the effective temperature, hence observed colour.  
This result has been subsequently questioned \citep[e.g.][]{girardi2011} with a number of studies supporting the idea that an age spread is responsible for the observed CMD features \citep{goudfrooij2011,  rubele2013}. \cite{yang2013} emphasized that the evolution of rotating models drastically depends on the parameters adopted in the description of the rotational mixing, so that the eMSTO can be explained --or not-- depending on which set of models are adopted. More recently \cite{bastiannieder2015} and  \cite{li2014Natur} excluded age spreads in intermediate age LMC clusters, by inspection of their sub giant branch morphology, but their analysis was challenged by a very careful study by \cite{goudfrooij2015}. At the same time, \cite{brandt2015} studied the turnoff spread expected on the basis of a rotation dispersion in the component stars based on the Geneva database created by C. Georgy and S. Ekstr\"om\footnote{http://obswww.unige.ch/Recherche/evoldb/index/} \citep{georgy2014popsint} and showed that, in the color magnitude diagram of the typical HST bands m$_{\rm F435W}$ versus color m$_{\rm F435W}$-m$_{\rm F814W}$, the turnoff area covered by turnoff stars increases, reaching a maximum towards 1--1.5~Gyr, which are in fact the typical ages of clusters with eMSTO. 

In this paper, we focus our attention of the LMC cluster NGC~1856, which is much younger than any other cluster with evident eMSTO. The eMSTO has been recently identified by \cite{correnti2015} and \cite{milone2015} who showed that this feature is consistent with an age spread of $\sim$80-150~Myr, although they do not exclude the rotational interpretation.\\
Interestingly, in addition to the eMSTO, this cluster exhibits a {\it split upper main sequence} which is clearly visible when the F336W filter is used (like in the m$_{\rm F555W}$ versus m$_{\rm F336W}$-m$_{\rm F555W}$ CMD). In a preliminary attempt dealing with models with age spread, \cite{milone2015} show that this feature may be explained by models in which the stars belonging to the cooler main sequence have a larger metallicity. While helium and light element variations are responsible for the MS splitting observed in the old galactic GCs, their effect on the double MS of NGC\, 1856 has been not yet investigated.
In this paper, by using the Geneva database, we show that the main cluster features can be interpreted in the framework of  the rotational hypothesis, and provide a possible explanation for the split MS.
 \section{Models}
\label{input}
The rotational interpretation of the extended turnoffs is linked to the initial distribution of the cluster stars' rotational velocities. Taking as an example the nominal age of 300~Myr attributed to NGC~1856 by \cite{milone2015}, the turnoff stars have  mass of about 3\msun, so the hydrogen core burning phase is convective.
The evolution of stars having a convective core in the main sequence is dramatically affected by rotation. In fact, the mechanism of chemical mixing associated with the transfer of angular momentum from the (more rapidly rotating) convective core to the stellar envelope provides fresh hydrogen to the burning core, and extends the main sequence phase. 
Fig.\ \ref{f1} illustrates the difference in the isochrone location for an age of 300~Myr and metallicity Z=0.006, having no rotation or a rotation equal to 0.5 the critical rotation rate \citep{georgy2013}.
The turnoff is more luminous by $\sim$0.3~dex. The masses in evolution at the turnoff are $\sim$2.9~\msun\ in the non rotating case, and  $\sim$~3.2\msun\  for  model rotating 0.5~\wcrit\ (central panel of Fig.~1). This example clearly illustrates how a range of rotational velocities in a population of coeval stars can be characterized by a broad range of turnoff luminosities and evolving masses, similar to those one would find in a cluster with non rotating stars spread over a range of ages. The hydrogen consumption and the luminosity evolution versus time are shown in  the right panel of Fig.~1,  for the 3\msun\ evolution. This panel makes clear that the time-luminosity evolution will depend on the detailed assumptions made concerning the mechanisms of chemical mixing.\footnote{These will be due both to the assumptions on rotational mixing, and to the form of overshooting of the convective core, adopted independently from rotation (that is, valid also in non rotating models). 
In the stellar models used for this study, the overshoot has been parametrized so that the convective core is extended by 0.1~$H_{\rm P}$ for stars more massive than 1.7~M$_\odot$, 0.05~$H_P$ for stars between 1.25 and 1.5~$M_\odot$ and none below. This calibration has been made for the rotating models to reproduce the width of Galactic open clusters (see Ekstrom et al. 2012). Non rotating models use the same calibration for the overshoot. Thus the differences seen in Fig.~1 results only from the effect of rotational mixing.}

\begin{figure}    
\centering{
\includegraphics[width=8.5cm]{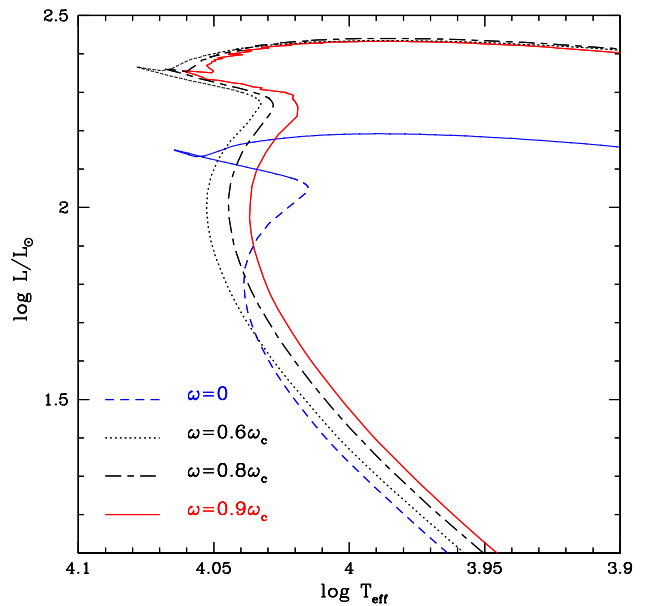}
}
\caption{Comparison of isochrones in the theoretical HR diagram, for age 350~Myr, Z=0.006 and different stellar angular velocity. Looking at the MSs, from left to right we have, in units of \wcrit, $\omega$=0 (dashed blue), 0.6 (dotted black), 0.8 (double dashed black) and 0.9 (full line red). } 
\label{f2} 
\end{figure}

It is therefore important to keep in mind that the detailed relation between turnoff spreads and rotation rates needed to model the observed CMD features depends on the specific parameters adopted in the stellar models used; a smaller rotation rate range would be needed if, for example, a higher chemical mixing efficiency than adopted in these specific models were assumed. 
In particular, in the Geneva database we use, the turnoff luminosity is not linearly related with the rotation velocity assumed for the stars, while the effect of rotation of the MS \teff\ is. We show this in Fig.~\ref{f2}, where isochrones of age t=350Myr are compared for $\omega/\omega_{\rm crit}$=0, 0.6, 0.8 and 0.9 in the theoretical plane. We see that a significantly cooler MS requires adopting a very large rotation. This effect is increased in the observational plane by the corrections due to limb and gravity darkening. \\

\section{The color magnitude diagram of NGC 1856: comparison with rotating models}
\label{cmd}
\begin{figure*}    
\centering{
\includegraphics[width=14cm]{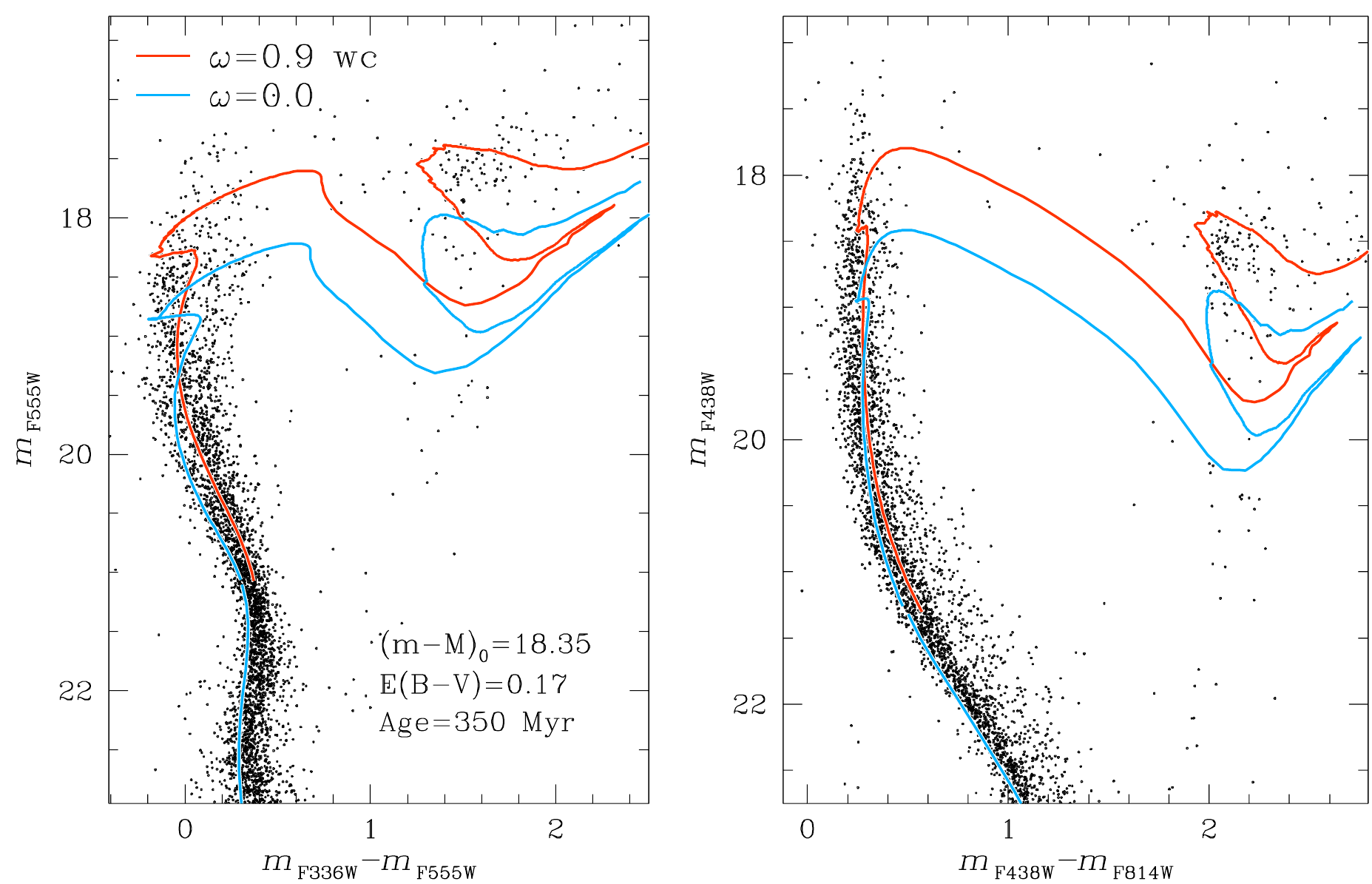}
}
\caption{Two isochrones of 350Myr for Z=0.006, used for the final simulation shown in Fig.~\ref{f4}, are  overimposed  to the data from Milone et al. (2015). The left panel shows the  m$_{F555W}$ vs m$_{\rm F336W}$-m$_{\rm F555W}$, and the right panel shows the m$_{F555W}$ vs m$_{\rm F438W}$-m$_{\rm F814W}$ color magnitude diagrams.  Two different value of angular velocities  ($\omega=0.9 \times \omega_{\rm crit}$ and $\omega=0$) are plotted. The non rotating isochrone is prolonged towards low masses using the models by Mowlavi et al. 2012. This allows to choose the value of distance modulus and reddening, labelled in the figure. The absorptions adopted in the different bands are  A$_{\rm 336W}$/E(B-V) =5.10, A$_{\rm 438W}$/E(B-V) =4.18, A$_{\rm 555W}$/E(B-V) =3.27, A$_{\rm 814W}$/E(B--V) =1.86 from Milone et al. 2015.} 
\label{f3} 
\end{figure*}

\begin{figure*}    
\centering{
\includegraphics[width=15cm]{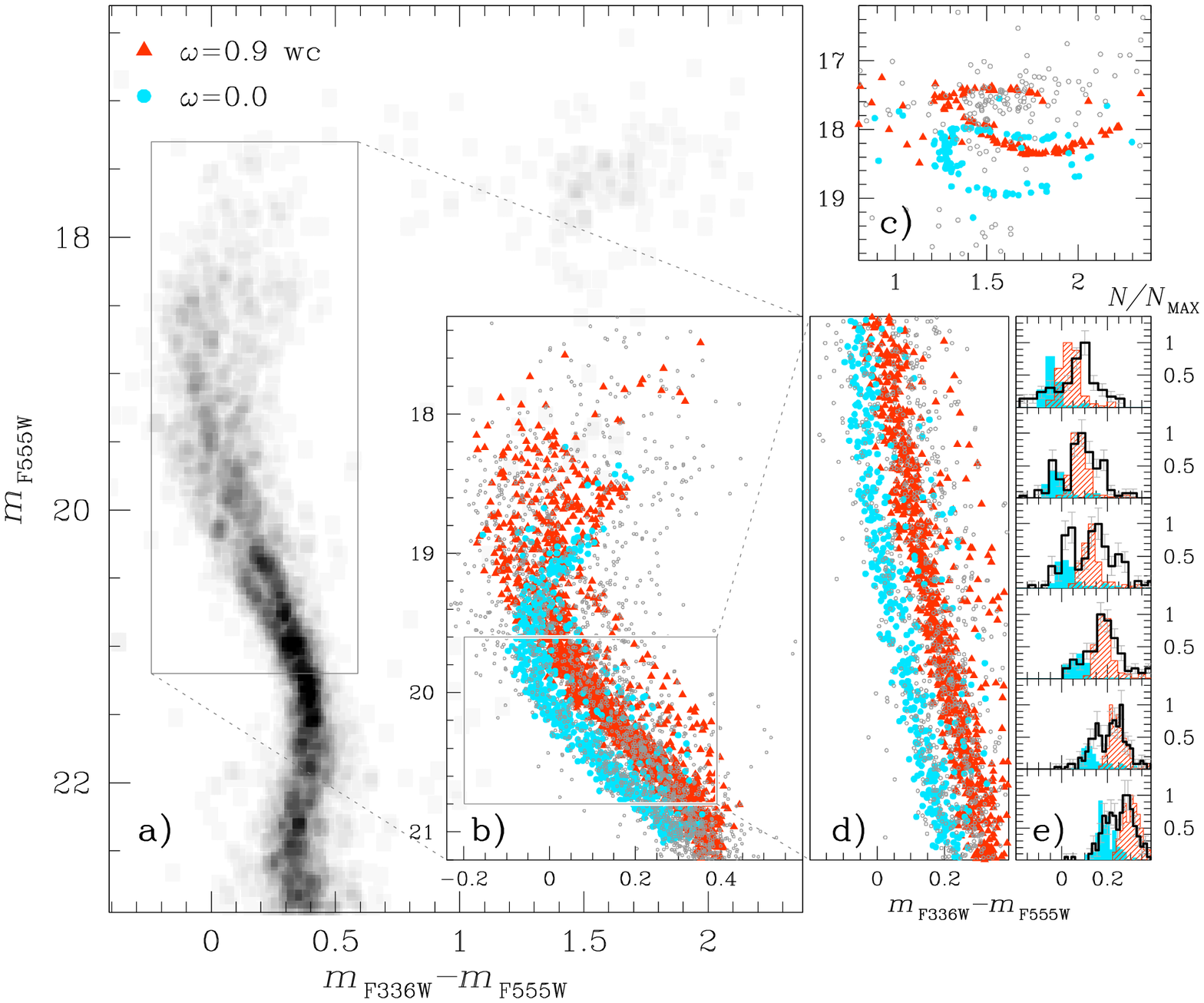}
}
\caption{ m$_{F555W}$ vs m$_{\rm F336W}$-m$_{\rm F555W}$ color magnitude diagram of cluster NGC1856 and its Hess diagram. Data (grey open circle) from Milone et al. (2015)  are compared with the simulations (filled triangles) described in detail in the text, at the luminosity of the turnoff (panel b); the red clump (c); and the upper part of MS (d). For this last case, the histogram of the $\Delta$ (m$_{\rm F336W}$-m$_{\rm F555W}$) distribution  for stars in six magnitude intervals is reported in panel e, after having verticalized the upper part of the main sequence delimited  by the gray boundaries in panel b). } 
\label{f4} 
\end{figure*}

Fig.~\ref{f3}  shows the CMD of the cluster NGC 1856 in m$_{\rm F555W}$ versus the m$_{\rm F336W}$-m$_{\rm F555W}$ color  observed by \cite{milone2015} (left panel) and m$_{\rm F438W}$\  versus m$_{\rm F438W}$-m$_{\rm F814W}$ color (right panel).
Fig.~\ref{f4}  shows the Hess diagram in the m$_{\rm F336W}$-m$_{\rm F555W}$ color. This is the first evidence of an extended main sequence turn-off ($\simgt$1mag) in a young cluster in the Magellanic Clouds as discussed by \cite{milone2015} and \cite{correnti2015}. 

In Fig.~\ref{f4} note also the split  of the upper MS in two component, which host $\sim$ 33$\%$ and 67$\%$  of MS stars in the red and blue part respectively. The detailed analysis of \cite{milone2015} has
demonstrated that both the split of MS and the eMSTO can not be due to photometric errors, differential reddening or unresolved binaries. The authors show that  if the stars in NGC1856 are chemically homogeneous and the eMSTO is due to a prolonged star formation, the star formation must have lasted about 150 Myr (not necessarily in a continuous way).  

As discussed in \S \ref{input} a range of stellar rotation velocities can mimic an age spread; here we show which distribution of rotational velocities is needed to reproduce the eMSTO and of the split of the upper MS.\\
In Fig.~\ref{f3} we show the comparison between data and rotating versus non--rotating stellar models.
 The isochrones, both for 350~Myr, are taken from the Geneva database  and  are calculated from the models by \cite{georgy2013}.
The metallicity we adopt is Z=0.006 that,  the metallicity, available on the database, closest to that  inferred from spectroscopy \citep{HZ2009} for the young populations of the Large Magellanic Cloud.\\
For the non rotating case we extend  the isochrones to the masses M$<$1.6\msun\ by means of the models published by \cite{mowlavi2012}. The extension  helps  to choose a suitable reddening and  distance modulus, which are labelled in Fig.\,\ref{f3}.
In order to reproduce the split in the upper MS, for the selected age, the population of rotating stars must be characterized by a rotation rate close to break-up  ($\omega \sim 0.9 \times \omega_{\rm crit}$) (Fig.\,\ref{f3}, left panel). Notice that the split of MS is less evident in  the color m$_{\rm F438W}$-m$_{\rm F814W}$ (Fig.\,\ref{f3}, right panel),  and the MSTO has a vertical structure. For this reason, UV photometry is crucial to shed light on the eMSTO phenomenon in NGC\,1856.
With this choice of rotation and age we then select from  the database the corresponding  synthetic photometry with a random viewing angle distribution, using the gravity darkening model by \cite{espinosa2011} and including  the limb darkening effect \citep{claret2000}. The data of the synthetic simulation were transformed into observational planes using  the model atmospheres by \cite{castelli2003}, convolved with the HST filter transmission curves.\\
In the simulation we take into account the combined photometry for  30\% of binaries, following the procedure outlined in \cite{milone2009}. Binaries are included both in the rotating and non rotating group. The mass function of the secondary stars is randomly extracted  from a Salpeter's like mass function, with a lower limit of 0.4\msun. Notice that only the photometric consequences of the presence of such binaries are monitored in the simulations (for the evolutionary consequences, see the discussion in \S \ref{sec:4.1}), and that the choice for the percentage of binaries is chosen based on the estimate by \cite{milone2015}.
\\
As shown in Fig.\,\ref{f4}, this  simulation describes well the eMSTO and the bimodality of the upper MS. \\
In other aspects the simulation is not completely satisfactory. In particular it fails to reproduce  the data after the end  of the central H-burning phase. We suggest that this may depend both on the parametrization of the overshoot in the Geneva models and on the way  in which the luminosity and temperature are corrected for the inclination  angle (see also discussion in \S \ref{sec:4.1}).\\
The simulation fails also to reproduce the red clump (panel c in Fig.\, \ref{f4}), but for a possible solution of this problem see \S \ref{sec:4.1}.

\section{Discussion}
\label{sec:disc}

\subsection{General adequacy of rotating models}
\label{sec:4.1}
The results shown in the panels of Fig.\,\ref{f4} rely on the Geneva database, and should be considered as a first attempt to explain the eMSTO and the split upper main sequence in NGC~1856 in terms of differences in the rotational velocities of  stellar subsets.
Noticeably, the Geneva database was built before the recent discoveries on NGC\,1856 and it is remarkable how the models do such a good job in reproducing its split main sequence.

There are two incomplete issues in the synthetic models: first, the red side of the eMSTO region is not well reproduced. It looks like the end of core hydrogen burning is not as sharply defined in reality as it occurs in the models we have used. This looks plausible, as the rotation and mixing issue may be subject to several minor second--order parameters which can not easily be taken into account in modeling. Second, the splitting of the main sequence is best reproduced  if we assign a very fast rotation (\wm=0.9~\wcrit) to the numerous sample (67\% of stars) defining the redder MS and the upper turnoff. This may depend on the way gravity and limb darkening are introduced in the models. 
This is the first adequate representation of the splitting, so it is a powerful hint that rotation is the reason for the CMD features, even if we should not take the very high rotation at face value. \\
Alternatives to the rotational interpretation can reside in different metallicity, or different specific elemental abundances for the two sides of the main sequence, but further modelling is required, and whether such differences exists should be checked by future spectroscopic observations.

The final problem is that the clump reproduction is quite modest. While the rotating simulation reproduces the luminosity location of the clump, there should be a second clump (less populated) about 1 magnitude dimmer, but the data do not show it. The clump evolution is a very delicate issue, the ratio of blue to red stars depending on the detailed treatment of convection and mixing. Consideration of non instantaneous mixing produces favors a long evolution in the blue side of the clump \citep[e.g.][]{venturacaststraka2005}, so it could describe better the clump here. 
Here again, notice that age differences produce a similar problem for the clump(s) location.
A possible solution of this problem occurs naturally if the slow-rotating stars are all binaries, as discussed in \S.~\ref{explain}. Observational data and theoretical considerations show that orbital periods between 4 and 500 days can be effective in producing slow stellar rotation \citep{abt2004}. As we are dealing with evolving masses $\sim$3\msun, orbital periods between $\sim$50 and 500 days, according to Kepler's law, would produce Roche lobe interaction and mass transfer from the primary star while it is evolving along the RGB. Shorter periods produce mass transfer at an earlier phase, during the crossing of the Hertzsprung gap. Thus the evolving star would not reach helium ignition as a single star would, and the low luminosity blue hook could result depopulated. In the end, the lack  or sparseness of a low luminosity clump might lend further support to the hypothesis of a rotational origin for the splitting of the turnoff.

A consequence of this interpretation is that the role of binaries in the  CMD should be reconsidered. The slowly rotating group should be made of 100\% binaries with periods between 4 and 500 days, while the rapidly rotating group would include both single stars and longer period binaries. The binaries clearly seen on the right of the observed main sequence(s) could be mostly belonging to the non rotating population. A full discussion of this problem goes beyond the purpose of this work, and is delayed to a following investigation.
 
\subsection{Why two populations?}
\label{explain}

Our analysis has shown that in order to reproduce the observed features of
the CMD of NGC 1856 it is necessary to have a large fraction of cluster
stars (2/3) retaining the large rotational velocity typical of the B and A
stars \citep[see e.g.][for A-type stars]{zorec2012}, while the remaining 1/3 of the stars must have slowed down early enough to prevent rotational mixing, and have evolved as standard non-rotating or slowly rotating models. 

The key question to address therefore concerns the mechanism and processes responsible for slowing down a large fraction of stars during the H--core burning evolution.

\cite{zorec2012} have gathered and analyzed the rotational velocities of fields A-type main-sequence stars (types B6 to F2). They found that stars below 2.5\msun\ show an unimodal velocity distribution displaying a small acceleration as the stars evolve along the MS. On the contrary, stars between 2.4 and 3.85\msun\ display a clear bimodal distribution with a slow velocity component (up to 20\% for the 2.8\msun\ bin). They found that the two peaks are clearly separated during the MS evolution but these peaks tend to bend near the turn-off. This dichotomy in velocity persists if close binary and chemically peculiar stars are included. The origin of the slow component is unknown but the authors make two hypotheses: (1) some stars have already lost all their angular momentum during the pre-MS due to magnetic braking or (2) in a fraction of undetected binaries, tidal interactions have slowed down the stars.
It should also be noted that a similar bimodality in the velocity distribution have been detected among early B-type stars by \cite{dufton2013}. 

\cite{abt2004}, in a study of the velocities of A and B-type binaries, found  that close binaries (period smaller than 4 days) are all synchronized and circularized. Binaries with period longer than 500 days display the same rotational properties as single stars. For binaries with period between 4 and 500 days, rotational velocity is significantly smaller than for single stars, with about one-third to two-third of their angular momentum being lost, presumably, by tidal interactions.

On the theoretical side, tides in binaries will be responsible to the synchronization and circularization of the orbit \citep{zahn1977}. For
low-mass stars with a convective envelope, the main mechanism is the viscous dissipation of the kinetic energy acting like an equilibrium tide. On the other hand, in more massive stars with a convective core, some low frequency modes of oscillation can be excited by the periodic tidal potential. The response to this excitation is called the dynamical tide \citep[see e.g.][]{Kopal1968}. With a simplified model, assuming  an uniform  rotation, \citet{zahn1977} obtains the following expression for the synchronization time:
\begin{equation}
\frac{1}{t_{\rm sync}} = 5 \times 2^{5/3}\left( \frac{GM}{R^3} \right)^{1/2} q^2 (1+q)^{5/6} \frac{MR^2}{I} E_2 \left( \frac{R}{a} \right)^{17/2}
\end{equation}
where $q$\ is the mass ratio of the binary, $E_2$\ is the parameter measuring the coupling between the tidal potential and the gravity mode which depends on the convective core mass ($\sim 5\times 10^{-8}$\ for a 3\msun).
\citet{zahn2008} shows that the synchronization time scales with $(I/MR^2)/E_2R^7$, which
  increases during the main sequence. Thus,  we can conclude that most of the
tidal interaction will take place around the ZAMS, so that binaries dispersion, if it occurs during the main sequence, will take place in systems already slowed-down.
Moreover, early-type binaries have their transition period (for circularization) observed through OGLE and MACHO survey well in agreement with the dynamical tides theory \citep{northzahn2003}. 
However an important fraction of binaries are also circularized even if the semi-major axis is larger. This may indicate that another more efficient mechanism is at play for them. One possibility will be that these binaries have undergone several episodes of resonance locking \citep[see][for the the effect on a 10\msun\ primary]{witte1999a, witte1999b} 

In summary, there are several ways to slow down a fraction of the cluster stars (see also the extended discussion in Dufton et al. 2013). If we look into more detail into the binary hypothesis, we should expect that most of the non--rotating stars hide a binary companion. In fact, binaries of periods in the range 4--500~days would be hard binaries in the core of this cluster, whose central dispersion velocity is $\sim$2.5~km/s (McLaughlin \& van der Marel 2006). Only binaries with period longer than $\sim10^3$yr would be ionized in the central regions. Notice that the cluster seems to host $\sim$30\% \citep{milone2015} of binaries, so this hypothesis is not unreasonable.  Further investigations may provide further insight into this problem.

\subsection{The case of multiple populations in old globular clusters}
We conclude our discussion with some comments and cautionary remarks about the connection between eMSTO clusters and multiple-population old GCs.

Evidence of multiple populations in old GCs comes from an extensive
number of photometric and spectroscopic studies showing that stars in
old GCs are characterized by a spread in the abundances in helium and
light elements (such as Na, O, Mg, Al). Photometric studies have
also shown the widespread presence of discrete groups with distinct
photometric properties corresponding to different chemical properties
\citep[see e.g.][]{piotto2015}. These are the key fingerprints of
multiple populations in old GCs and provide the main constraints for
any theoretical effort aimed at modeling the formation history of
these systems and shedding light on the source of processed gas
composing the second-generation populations observed in globular
clusters.  

While the study of young and intermediate-age clusters (hereafter
YICs) might provide some clues to the answers to the many questions
raised by the discovery of multiple populations in old GCs it is
important to exercise much caution in connecting YICs and old GCs for
this purpose and in drawing conclusions about different theoretical
scenarios for the formation of old GCs from observations of YICs. 

Some specific issues to consider are the following.

1) It is not clear whether the YICs studied so far actually host
multiple populations at all. As mentioned in the Introduction, for
example, a spectroscopic study of one  of the eMSTO clusters (NGC
1806) does not reveal the typical chemical fingerprints of multiple
populations found in old GCs. Photometric signatures which might be
ascribed to extended
star formation history in YICs, such as those discussed in this and other
works, have possible alternative explanations unrelated to the
presence of multiple stellar populations with different chemical
properties. It is important to realize that  spectroscopic
characterization of YIC  aimed at shedding light on the presence (or lack of) the chemical patterns typical of multiple populations is desirable to make a solid connection between YIC and old multiple-populations GCs even if
strong evidence of an age spread in YIC were to be established.
Although, as remarked in \S.\ref{Intro}, many additional spectroscopic studies are necessary to
clarify this issue, it is possible that the study of these clusters might be irrelevant to shed light on the formation history of multiple-population old GCs. 

2) While eMSTO, as also argued in this paper, might indeed be explained
without relying to extended periods of star formation (and eMSTO
clusters might not host chemically different multiple populations), 
studies ruling out
star formation episodes extending for a few hundred Myr \citep[see, e.g.][]{bastiansilvavilla2013, bastian2013a}
have been presented as evidence against the AGB scenario of
multiple-population formation. We point out here that in all the
studies \citep[see e.g.][]{dercole2008, dercole2010, dercole2012} aimed at modeling
the chemical properties of multiple-population old GCs  using AGB
ejecta the epoch of second-generation formation is limited to a time
interval between about  30 and 100 Myr with a significant fraction of
second generation stars forming in the early phase of this
period. While in a few clusters second-generation star formation might
have extended beyond \citep[see e.g.][]{dantona2011wcen} the current
observational constraints from spectroscopic studies ---in particular the lack of C+N+O increase
in the second generation stars of most clusters--- do not, in
general,  allow such an extended period of second-generation star
formation. In addition to urging caution in connecting YICs and old
GCs, it is important to realize that lack of evidence of such an
extended age spread can not be used as evidence against AGB models. 

3) Finally we emphasize the importance of properly considering the
differences  between the structural properties of YIC and old
GCs. The current properties of old GCs have been affected by early and
long-term evolutionary processes (e.g. gas expulsion, mass loss due to
stellar evolution, two-body relaxation, tidal
interactions). Reconstructing their structure and mass at the time of
second-generation formation in order to identify the threshold in the
structural properties allowing the formation of multiple populations
is not a trivial task. Identifying young clusters with present masses
similar to the current masses of old multiple-population clusters to test
multiple-population formation scenarios, without considering the
complications of properly reconstructing the 
evolutionary history of old GCs, may easily lead to erroneous conclusions. For
example the current mass of NGC 1856 is about $10^5~M_{\odot}$ \citep{mclaughlin2005},
similar to the current mass of some old GCs hosting multiple populations. Its
structural properties are such the central escape speed is about 10
km/s \citep{mclaughlin2005}. Detailed models would be necessary to understand whether
initial/earlier properties allowing the formation of a significant fraction of second-generation stars are among those
evolving into NGC 1856's current properties ---see \cite{correnti2015} for an attempt to address this issue; see also \cite{goudfrooij2014}. 
However the current structural properties, if sufficiently similar to those of the cluster at earlier stages of its life, would indicate that this cluster is not capable to retain a significant fraction of the ejecta of AGB or other possible sources of processed gas, lend support to the interpretation of the eMSTO in terms of different rotational velocities, and provide an example of the caution necessary in using these stellar systems to draw any
conclusion about multiple population formation models.

\section*{Acknowledgments}
We acknowledge a vast and proficuous use of the interactive tool of Geneva stellar models at the web address 
http://obswww.unige.ch/Recherche/evoldb/index/, a facility created and maintained by C. Georgy and S. Ekstr\"om. We thank the anonymous referee for a detailed and useful report.

F.D'A. acknowledges support from PRIN INAF 2014 (principal investigator S. Cassisi). M.D.C. and P.V. acknowledge support from INAF-OAR. 
T.D. acknowledges support from the UE Program (FP7/2007/2013) under grant agreement number 267251 of Astronomy Fellowships in Italy (ASTROFit). 
A.P.M. acknowledges support by the Australian Research Council through Discovery Early Career Researcher Award DE150101816. E.V. acknowledges support from grant NASANNX13AF45G.  







\bsp	
\label{lastpage}
\end{document}